\documentclass[%
 reprint,
superscriptaddress,
 amsmath,amssymb,
 aps,
]{revtex4-1}

\usepackage{graphicx}% Include figure files
\usepackage{dcolumn}% Align table columns on decimal point
\usepackage{bm}% bold math
\begin{document}

\title{Generation and applications of an ultrahigh-fidelity four-photon Greenberger-Horne-Zeilinger state}

\author{Chao Zhang}
\affiliation{Key Laboratory of Quantum Information, University of Science and Technology of China, CAS, Hefei, 230026, China}
\affiliation{Synergetic Innovation Center of Quantum Information and Quantum Physics, University of Science and Technology of China, Hefei, 230026, P.R. China}

\author{Yun-Feng Huang}
\email{hyf@ustc.edu.cn}
\affiliation{Key Laboratory of Quantum Information, University of Science and Technology of China, CAS, Hefei, 230026, China}
\affiliation{Synergetic Innovation Center of Quantum Information and Quantum Physics, University of Science and Technology of China, Hefei, 230026, P.R. China}

\author{Chengjie Zhang}
\affiliation{College of Physics, Optoelectronics and Energy, Soochow University, Suzhou, 215006, China}
\affiliation{Centre for Quantum Technologies, National University of Singapore, 3 Science Drive 2, Singapore 117543, Singapore}

\author{Jian Wang}
\affiliation{Key Laboratory of Quantum Information, University of Science and Technology of China, CAS, Hefei, 230026, China}
\affiliation{Synergetic Innovation Center of Quantum Information and Quantum Physics, University of Science and Technology of China, Hefei, 230026, P.R. China}

\author{Bi-Heng Liu}
\affiliation{Key Laboratory of Quantum Information, University of Science and Technology of China, CAS, Hefei, 230026, China}
\affiliation{Synergetic Innovation Center of Quantum Information and Quantum Physics, University of Science and Technology of China, Hefei, 230026, P.R. China}

\author{Chuan-Feng Li}
\email{cfli@ustc.edu.cn}
\affiliation{Key Laboratory of Quantum Information, University of Science and Technology of China, CAS, Hefei, 230026, China}
\affiliation{Synergetic Innovation Center of Quantum Information and Quantum Physics, University of Science and Technology of China, Hefei, 230026, P.R. China}

\author{Guang-Can Guo}
\affiliation{Key Laboratory of Quantum Information, University of Science and Technology of China, CAS, Hefei, 230026, China}
\affiliation{Synergetic Innovation Center of Quantum Information and Quantum Physics, University of Science and Technology of China, Hefei, 230026, P.R. China}

\date{\today}

\begin{abstract} High-quality entangled photon pairs generated via spontaneous parametric down-conversion have made great contributions to the modern quantum information science and the fundamental tests of quantum mechanics. However, the quality of the entangled states decreases sharply when moving from biphoton to multiphoton experiments, mainly due to the lack of interactions between photons. Here, for the first time, we generate a four-photon Greenberger-Horne-Zeilinger state with a fidelity of $98\%$, which is even comparable to the best fidelity of biphoton entangled states. Thus, it enables us to demonstrate an ultrahigh-fidelity entanglement swapping---the key ingredient in various quantum information tasks. Our results push the fidelity of multiphoton entanglement generation to a new level and would be useful in some demanding tasks, e.g., we successfully demonstrate the genuine multipartite nonlocality of the observed state in the nonsignaling scenario by violating a novel Hardy-like inequality, which requires very high state-fidelity. \end{abstract}

\pacs{03.65.Ta, 03.65.Ud, 42.50.Dv, 42.50.Xa}

\maketitle

\section{Introduction}

Photons are promising candidates for quantum information processing \cite{nielsen2010quantum}, due to their weak interaction to environment and easy single-qubit operations. In photonic quantum information processing, preparing high-quality entangled states of photons plays a key role. The reason is that photons lack of interactions, while photonic quantum information protocols usually can be realized with off-line-generated entangled photons, such as the one-way quantum computation \cite{PhysRevLett.86.5188,PhysRevLett.86.910}, and the quantum teleportation \cite{PhysRevLett.70.1895}. Today, biphoton entanglement generation has become general in laboratory. The most convenient way arises from the spontaneous parametric down-conversion (SPDC) processes in nonlinear crystals. Extremely high-fidelity biphoton Bell states have been prepared via SPDC processes recently \cite{PhysRevLett.115.180408,PhysRevX.5.041052} (up to $99.9\%$ in Ref. \cite{PhysRevLett.115.180408}). Also, high-quality entangled photon pairs have been used for the loophole-free test of local realism \cite{PhysRevLett.115.250401,PhysRevLett.115.250402}.

However, the state fidelity of three or more entangled photons is still at a much lower level \cite{HamelShalmHuebelEtAl2014,PhysRevLett.86.4435,jennewein2009performing}, mainly due to the difficulty of entangling photons from independent sources. Entangling independent photons is also the key ingredient in various quantum information tasks, such as entanglement swapping \cite{HalderBeveratosGisinEtAl2007} and teleportation. Thus, moving from biphoton to multiphoton entanglement means much more than a simple expansion.

In this paper, we experimentally generate an ultrahigh-fidelity four-photon Greenberger-Horne-Zeilinger (GHZ) state, with only 20 mW pump power, using our new ``sandwichlike" Einstein-Podolsky-Rosen (EPR) source \cite{PhysRevLett.115.260402}. This EPR source is based on the beamlike type-II phase matching \cite{Takeuchi:01,kurtsiefer2001generation}, and is rather suitable for multiphoton entanglement generation. It achieves high brightness, high fidelity and high collection efficiency at the same time. The high collection efficiency and high-fidelity enables high-performance multiphoton experiments. And the high brightness greatly reduces the power requirement for the pump laser employed in multiphoton experiments. With the high-fidelity GHZ state, we further demonstrate a high-performance entanglement swapping operation. We also test a demanding Hardy-like inequality in the nonsignaling senario, which can detect genuine n-partite nonlocality using only 2n local measurement settings, but require rather high state fidelity.

\section{Generation of the four photon GHZ state.} SPDC in nonlinear crystals provides a convenient way to produce entangled photon pairs. According to the phase-matching type, there are two kinds of widely used entanglement sources. The type-I entanglement source \cite{PhysRevA.60.R773} uses a two-crystal geometry (closely placed, relatively thin, identically cut). The down-converted photons from the two crystals almost emit as one cone centered at the pump beam due to the small thicknesses of the two crystals. The optic axes of the two crystals lie in the vertical and horizontal plane, respectively. So, a $45^\circ$-polarized photon in the pump laser will be down-converted equally in either crystal. And the down-converted photon pairs from different crystals have orthogonal polarizations. When choosing two emitting directions from the cone symmetrically (about the pump beam), the collected photon pairs are in the polarization entangled state. In the type-II entanglement source \cite{PhysRevLett.75.4337}, the down-converted photons emit into two crossed cones, and have orthogonal polarizations. So, when photon pairs are collected along the two intersecting directions, they are in the Bell state. However, in both cases, most of the down-conversion fields are wasted, so the brightness of these sources are greatly limited.

To overcome this shortcoming, the beamlike type-II phase matching is designed and implemented \cite{Takeuchi:01,kurtsiefer2001generation}, in which the down-conversion fields shrink into two separate beams. This is accomplished by decreasing the phase-matching angle from the situation of two crossed cones. Each beam has a Gaussian-like intensity distribution and a small divergence angle, thus all down-conversion fields can be easily collected. With a two-crystal geometry, beamlike entangled photon pairs source has been designed \cite{PhysRevA.68.013804} and implemented \cite{Niu:08}. In a most recent work \cite{PhysRevLett.115.260402}, we further design and implement a new ``sandwichlike" EPR source (see Appendix) under the beamlike phase matching condition. The key idea is to make all ordinary (o) photons emitting in one path, while all extraordinary (e) photons emitting in the other path, as shown in Fig. 1. And the nonlinear crystals used for SPDC processes here are $\beta$-barium borate (BBO) crystals. Due to the high brightness and high quality of this source, we choose it to generate our four-photon GHZ state.

%%%%%%%%%%%% FIGURE 1 %%%%%%%%%%%%%%%%%%%%%%%%%%%%%%%%%%
\begin{figure}[tb]
\centering
\includegraphics[width=0.4\textwidth]{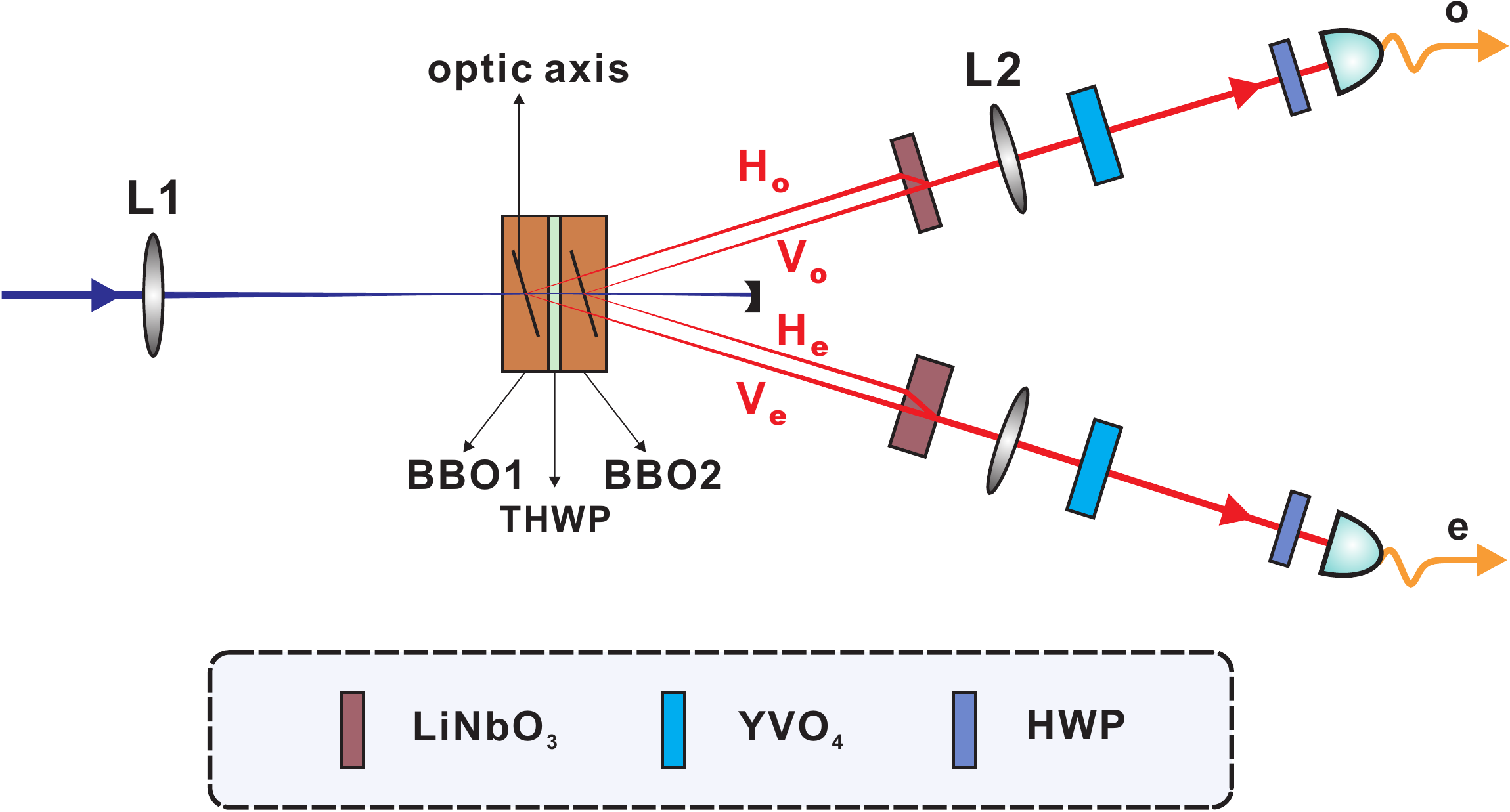}
\caption{\label{Fig:1}Experimental setup for our ``sandwichlike" EPR source. The thicknesses of the LiNbO$_3$ crystals are different in the two paths, which is due to the walk-off effect of the pump beam and the down-converted beams in the BBO crystals. The half-wave plates (HWPs) before the fiber couplers are used for polarization alignment.}  \end{figure}
%%%%%%%%%%%%%%%%%%%%%%%%%%%%%%%%%%%%%%%%%%%%%%%%%%%%%%%

Next, we show how to prepare the high-fidelity four-photon GHZ state from two sandwichlike EPR sources. The key requirement is the ability to entangle photons from independent sources. Due to the lack of interaction between photons, the general method is using the measurement induced nonlinearity \cite{KnillLaflammeMilburn2001}, e.g., using a projection operator to project the product state onto the entangled state with a certain probability. The seminal work \cite{zukowski1995entangling} gives us a feasible scheme to make it into practice. The key idea is to use pulsed pump laser and relatively narrow-band filters for interfering photons. Here we follow their scheme.

The experimental setup is shown in Fig. 2. The ultraviolet pump laser for each EPR source is generated in the same way as described in the Appendix, except that we tune the pulse duration of the mode-locked Ti:Sapphire laser to 90 fs for better multiphoton interference visibility. And two sandwichlike BBO strategies are employed to generate the entangled photon pairs 1, 2 and 3, 4. Thus the input state is
\begin{equation}
\label{eq:inputstate}
|\psi^{in}\rangle=\frac{1}{2}\left(|HH\rangle+|VV\rangle\right)_{12}\otimes\left(|HH\rangle+|VV\rangle\right)_{34}
\end{equation}
Here, $|H\rangle\ (|V\rangle)$ denotes the horizontal (vertical) polarization state of the photons. All photons are coupled into single-mode fibers for spatial filtering. The two e-photons (2 and 3) are directed to the PBS for the Hong-Ou-Mandel (HOM) interference. The PBS acts as a parity check gate $(|HH\rangle\langle HH|+|VV\rangle\langle VV|)$, such that the two interfering photons can emit into two different output ports only if they have the same polarization. Thus, when we postselect the cases that there is one and only one photon in each output port (1, $2^\prime$, $3^\prime$, 4), the input state is projected onto the four-photon GHZ state $|G_4\rangle=\frac{1}{\sqrt{2}}\left(|H\rangle^{\otimes4}+|V\rangle^{\otimes4}\right)$. Then, each photon passes through a narrow-band filter for spectral filtering. Here, to achieve a good trade-off between better collection efficiency and higher multiphoton interference visibility, we choose 2 nm and 3 nm FWHM filters for e- and o-photons, respectively.

%%%%%%%%%%%% FIGURE 2 %%%%%%%%%%%%%%%%%%%%%%%%%%%%%%%%%%
\begin{figure}[tb]
\centering
\includegraphics[width=0.4\textwidth]{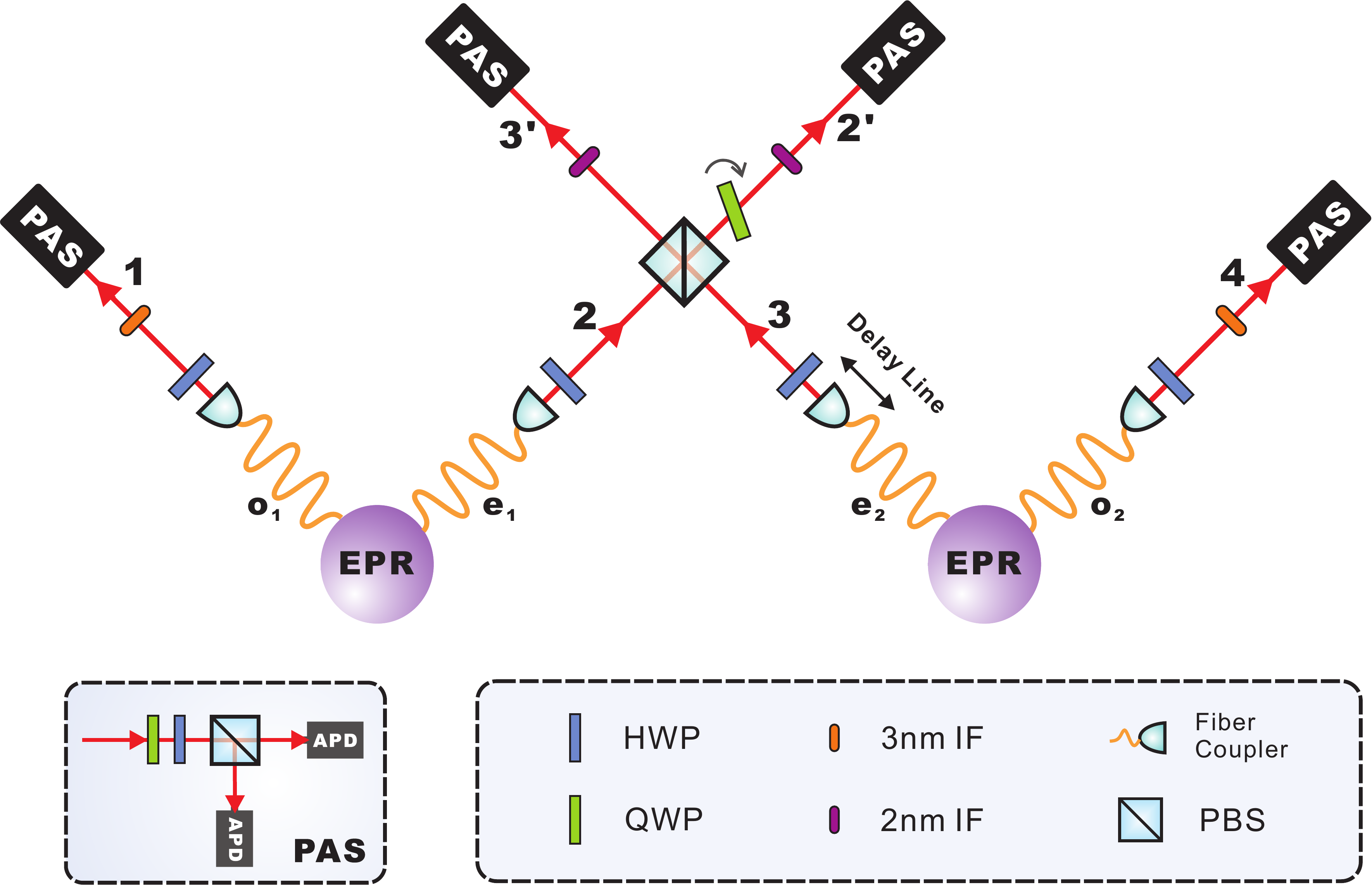}
\caption{\label{Fig:2}Experimental setup to generate the four-photon GHZ state. The abbreviations of the components are HWP, half-wave plate; QWP, quarter-wave plate; IF, interference filter; PBS, polarizing beam splitter; PAS, polarization analyzing system; APD, avalanche photodiode detector. The two EPR sources generated two pairs of entangled photons 1, 2 and 3, 4. The HWPs after the fiber couplers are used for polarization alignment. The two e-photons are directed to overlap on the central PBS, which can project the input state onto the four-photon GHZ state. One of the fiber couplers is mounted on a translation stage to finely adjust the arriving time of the interfering photons. The tiltable QWP after the PBS is used for tuning the relative phase between the two terms in the GHZ state. Each photon passes though a narrow-band filter for spectral selection and then enters the final PAS. The inset shows the details of the PAS, which consists of one motorized QWP, one motorized HWP, one PBS and two APDs. }  \end{figure}
%%%%%%%%%%%%%%%%%%%%%%%%%%%%%%%%%%%%%%%%%%%%%%%%%%%%%%%

To characterize the observed state, we use both the entanglement witness and quantum state tomography (QST). In the measurement part, each photon is detected by a polarization analyzing system (PAS), which can measure it in any desired polarization basis with a high alignment accuracy. The entanglement witness for the four-photon GHZ state \cite{PhysRevA.76.030305} is
\begin{equation}
\label{eq:witness}
W_{G_4}=\frac{I}{2}-|G_4\rangle\langle G_4|=\frac{I}{2}-\frac{1}{2}A-\frac{1}{4}\sum_{k=0}^3(-1)^kM_k
\end{equation}
where $A=|H\rangle^{\otimes4}\langle H|^{\otimes4}+|V\rangle^{\otimes4}\langle V|^{\otimes4}$ and $M _{k}=[\cos (\frac{k\pi }{4})\sigma _{x}+\sin (\frac{k\pi }{4})\sigma _{y}]^{\otimes4}, (k=0,1,2,3)$ are local measurement operators. In order to suppress the higher-order photon pair emission noise, we use a very low pump power of 20 mW and still obtain a moderate fourfold coincidence counting rate of 0.42 per second. The measurement result is shown in Fig. 3, which yields $\langle W_{G_4}\rangle=-0.4810\pm0.0023$, thus the state fidelity can be deduced as $F=Tr(\left\vert G_{4}\right\rangle \left\langle G_{4}\right\vert \rho_{exp} )=\frac{1}{2}-Tr(W_{G_{4}}\rho_{exp} )=0.9810\pm0.0023$, where $\rho_{exp}$ is the density matrix of the experimentally prepared state. On the other hand, we use the over-complete QST \cite{KaltenbaekLavoieZengEtAl2010,PhysRevLett.95.210504} to estimate the observed state, which needs 81 joint measurement settings. The total measurement time is over 6 hours. Then we use the maximum likelihood approach to reconstruct the density matrix $\rho_{exp}$. Figure 4 shows the real and imaginary parts of $\rho_{exp}$. The state fidelity is calculated to be $F=\langle G_4|\rho_{exp}|G_4\rangle=0.9794\pm0.0040$, which agrees with the above result very well.

%%%%%%%%%%%% FIGURE 3 %%%%%%%%%%%%%%%%%%%%%%%%%%%%%%%%%%
\begin{figure}[tb]
\centering
\includegraphics[width=0.5\textwidth]{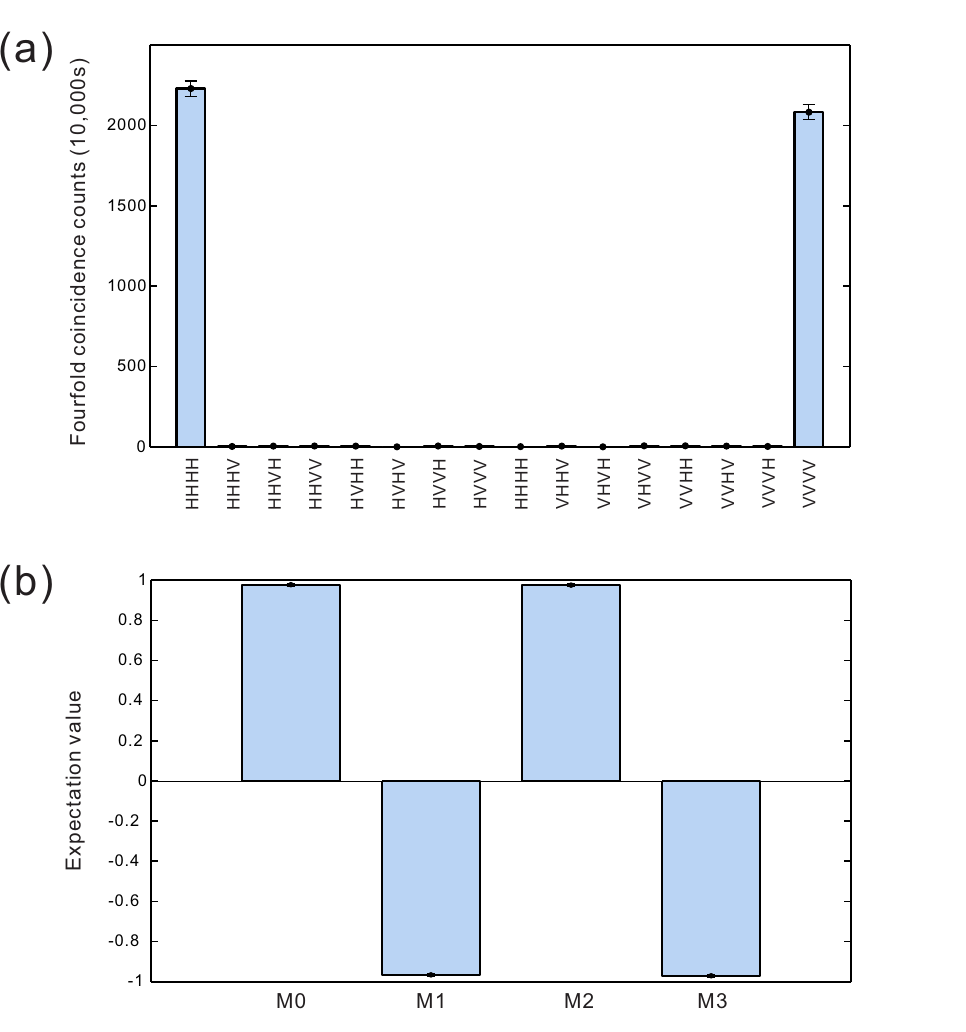}
\caption{\label{Fig:3}Measurement result of the witness. (a) Fourfold coincidence counts measured in the H/V basis. The data collection time is 10,000 s. The expectation value of A is just equal to the proportion of the two correct terms, which is $0.9897\pm0.0015$. (b) The expectation values of $M_k$, which yield an average value of $\frac{1}{4}\sum_{k=0}^3(-1)^k\langle M_k\rangle=0.9722\pm0.0041$. The data collection time is 2000 s for each measurement settings.  }  \end{figure}
%%%%%%%%%%%%%%%%%%%%%%%%%%%%%%%%%%%%%%%%%%%%%%%%%%%%%%%

%%%%%%%%%%%% FIGURE 4 %%%%%%%%%%%%%%%%%%%%%%%%%%%%%%%%%%
\begin{figure}[tb]
\centering
\includegraphics[width=0.5\textwidth]{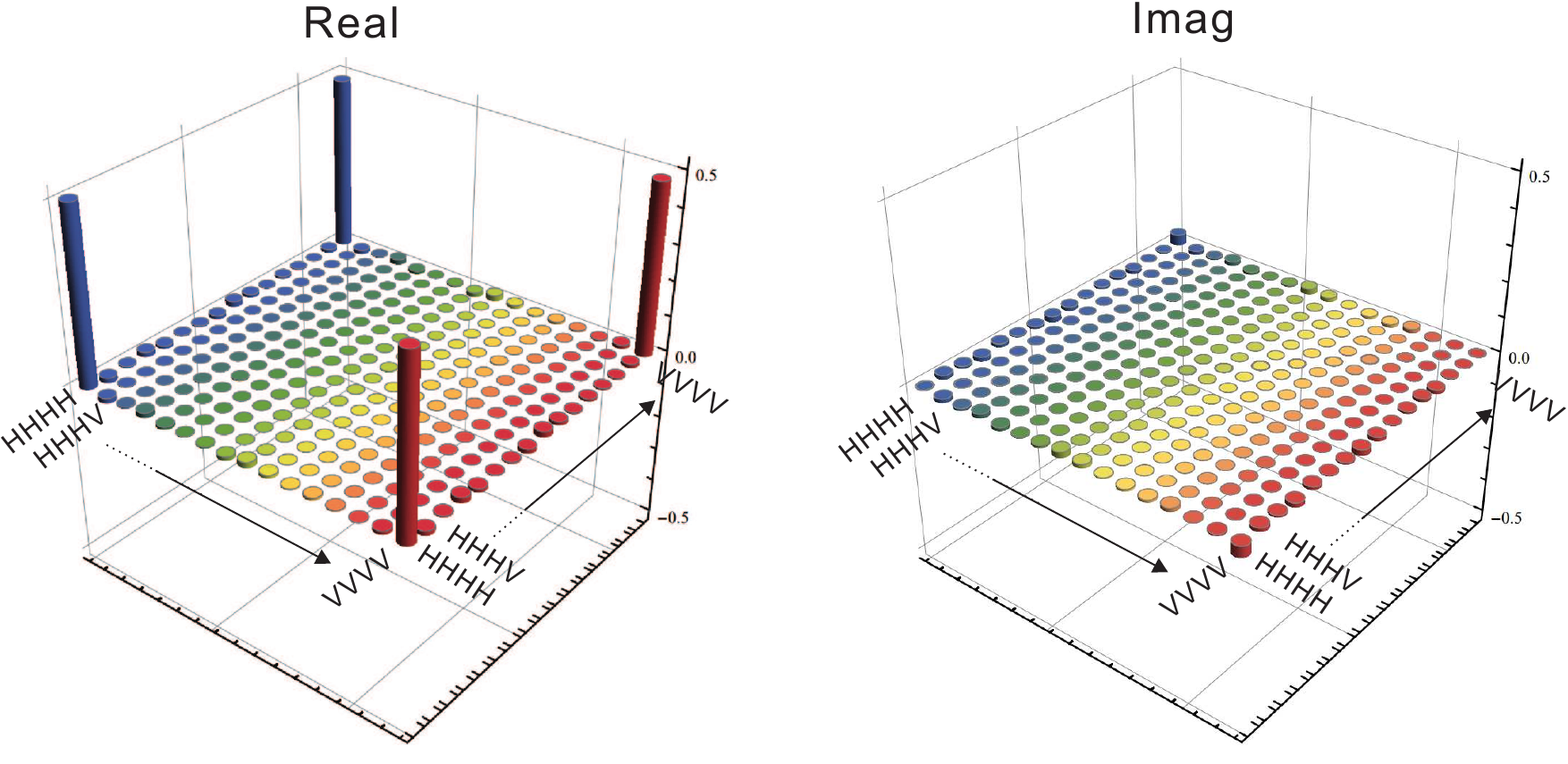}
\caption{\label{Fig:4}The real (left) and imaginary (right) parts of the density matrix $\rho_{exp}$, which is reconstructed by the maximum likelihood method from the recorded data. The height of the largest incorrect bar in the two pictures is only 0.03.}  \end{figure}
%%%%%%%%%%%%%%%%%%%%%%%%%%%%%%%%%%%%%%%%%%%%%%%%%%%%%%%

\section{Entanglement swapping with the sandwichlike EPR source.} Our setup can also be used to demonstrate entanglement swapping, the realization of which has rather profound implications. First, it is the key process in the quantum repeater \cite{RevModPhys.83.33}, which requires exchanging entanglement between different photon pairs. Second, in the distributed quantum computation \cite{grover1997quantum}, entanglement swapping allows us to connect distant nodes of a large network. Furthermore, it can also be viewed as teleportation of a genuinely unknown state \cite{PhysRevLett.70.1895}.

The principle of entanglement swapping can be understood by rewriting the product state in Eq. (\ref{eq:inputstate}) in the following way
\begin{eqnarray}
\label{eq:decompos}
|\psi^{in}\rangle = \frac{1}{2}(|\psi^+\rangle_{14}|\psi^+\rangle_{23}+|\psi^-\rangle_{14}|\psi^-\rangle_{23}\nonumber\\ +|\phi^+\rangle_{14}|\phi^+\rangle_{23}+|\phi^-\rangle_{14}|\phi^-\rangle_{23})
\end{eqnarray}
where $|\psi^{\pm}\rangle=\frac{1}{\sqrt{2}}(|HV\rangle\pm|VH\rangle)$, $|\phi^{\pm}\rangle=\frac{1}{\sqrt{2}}(|HH\rangle\pm|VV\rangle)$ are Bell states.
When we project the two independent photons 2 and 3 onto one of the four Bell states, the other two photons 1 and 4 will simultaneously collapse to the same Bell state, although they have never been able to interact with each other. In practice, the Bell state measurement can be realized by a PBS with two PASs in the output ports. The PASs are set to measure polarizations in the $+/-$ basis. For $|\phi^+\rangle$, the measurement result will be $|++\rangle$ or $|--\rangle$; for $|\phi^-\rangle$, the measurement result will be $|+-\rangle$ or $|-+\rangle$. For $|\psi^+\rangle$ and $|\psi^-\rangle$, the two photons end up at the same detector and with identical polarizations due to the HOM interference. Thus, we can only discriminate $|\phi^+\rangle$ and $|\phi^-\rangle$ here. If we want to discriminate the other two Bell states, we can insert a $45^\circ$ half-wave plate (HWP) in one of the input port of the PBS.

Since we have used the over-complete set of bases in QST to characterize the GHZ state as described hereinbefore, we can also use these data to demonstrate entanglement swapping, e.g., by picking out the data that photons 2 and 3 are measured in the +/- basis. Figure 5 shows the tomographic result for photons 1 and 4 when the photons 2 and 3 are projected onto $|\phi^+\rangle$ or $|\phi^-\rangle$. We can calculate the average swapping fidelity to be $0.977\pm0.01$, which, to our best knowledge, is the highest value ever reported. Here we just assume that the input EPR states are perfect, and calculate the fidelity between the state of photons 1, 4 and the corresponding Bell state.

%%%%%%%%%%%% FIGURE 5 %%%%%%%%%%%%%%%%%%%%%%%%%%%%%%%%%%
\begin{figure}[tb]
\centering
\includegraphics[width=0.5\textwidth]{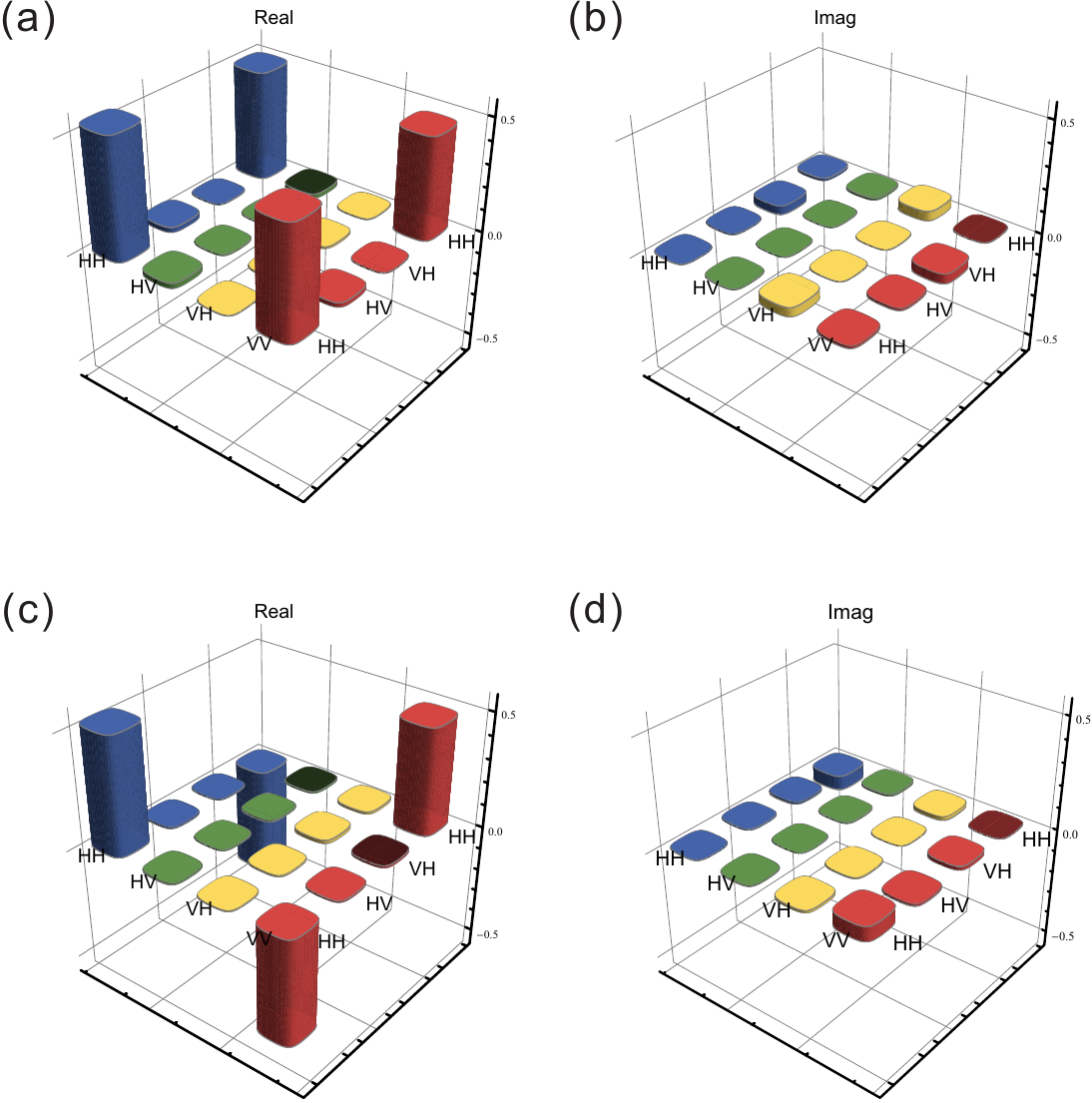}
\caption{\label{Fig:5}Real (left) and imaginary (right) parts of the reconstructed density matrix of photons 1, 4. (a), (b) The state of photons 2, 3 are projected on $|\phi^+\rangle$, the fidelity of the state of photons 1, 4 corresponding to $|\phi^+\rangle$ is $0.979\pm0.012$. (c), (d) The photons 2 and 3 are projected on $|\phi^-\rangle$, the fidelity of the state of photons 1, 4 corresponding to $|\phi^-\rangle$ is $0.975\pm0.016$. }  \end{figure}
%%%%%%%%%%%%%%%%%%%%%%%%%%%%%%%%%%%%%%%%%%%%%%%%%%%%%%%

\section{Testing the Hardy-like inequality.} High-fidelity multipartite entangled states are not only important for various quantum information protocols, but also useful in the fundamental test of quantum mechanics. Here, using the observed four-photon GHZ state, we experimentally test a novel Hardy-like inequality introduced in \cite{PhysRevLett.112.140404}, which can detect genuine multipartite nonlocality in the nonsignaling scenario. The inequality has some significant features. First, compared with the standard Svetlichny inequality \cite{PhysRevD.35.3066}, it needs only 2n joint measurement settings for n parties, which reduces much experimental effort (the Svetlichny inequality requires $2^n$ joint measurement settings). Second, it can prove that all entangled permutation symmetric pure states are genuine multipartite nonlocal. In the following, we can see that a high state-fidelity is necessary to violate the inequality.

\begin{table}[tb]
\renewcommand\arraystretch{1.3}
\centering
\caption{\label{Table:1} Measurement results of the Hardy-like inequality. The middle column lists the theoretical expectation values of the ideal GHZ state, for each measurement setting. In order to obtain a small error bar, the data collection time is 8 hours for each of the first two settings and 4 hours for each of the other settings, and we get totally 58,000 fourfold coincidence counts. The final result shows a violation of the inequality by more than 4 standard deviations. Note that all the raw data in each measurement settings have been corrected for the different detection efficiencies of the two APDs in each PAS. The error bars are deduced from the raw data and poisson counting statistics. }

\begin{tabular}{c c c} \hline
\ \ \ Settings\ \ \ &\ \ \ Theory\ \ \ &\ \ \ \ \ Experiment\ \ \ \ \ \\\hline
$\langle a_1aaa\rangle$            &$0.0479$     &$0.0501\pm0.0020$ \\
$\langle b_1aaa\rangle$            &$0.0131$     &$0.0160\pm0.0012$ \\
$\langle a_1baa\rangle$            &$0.0029$     &$0.0035\pm0.0008$ \\
$\langle a_1aba\rangle$            &$0.0029$     &$0.0048\pm0.0009$ \\
$\langle a_1aab\rangle$            &$0.0029$     &$0.0037\pm0.0008$ \\
$\langle \overline{b}_1\overline{b}aa\rangle$            &$0.0018$     &$0.0044\pm0.0009$ \\
$\langle \overline{b}_1a\overline{b}a\rangle$            &$0.0018$     &$0.0025\pm0.0007$ \\
$\langle \overline{b}_1aa\overline{b}\rangle$            &$0.0018$     &$0.0014\pm0.0005$ \\\hline
$I$&0.0209&$0.0138\pm0.0030$\\\hline
\end{tabular}
\end{table}

For $n=4$ case, the inequality has the following form:
\begin{eqnarray}
\label{eq:decomposition}
I&=&\langle a_1a_2a_3a_4\rangle-\langle b_1a_2a_3a_4\rangle-\langle a_1b_2a_3a_4\rangle\nonumber\\
&-&\langle a_1a_2b_3a_4\rangle-\langle a_1a_2a_3b_4\rangle-\langle \overline{b}_1\overline{b}_2a_3a_4\rangle\nonumber\\
&-&\langle \overline{b}_1a_2\overline{b}_3a_4\rangle-\langle \overline{b}_1a_2a_3\overline{b}_4\rangle\leq0.
\end{eqnarray}
Here $a_i$ and $b_i$ are binary valued observables (with outcomes 0 and 1) for the i-th party, and $\overline{b}_i=1-b_i$ has opposite outcomes compared with $b_i$. In our case, to demonstrate the genuine multipartite nonlocality of the GHZ state by violating this inequality, we need to find suitable dichotomous projection operators $a_i$ and $b_i$ for the i-th party, and $\overline{b}_i$ is orthogonal to $b_i$ ($\overline{b}_i=I-b_i$). To do this, we look for the projection measurements by numerical search. For simplicity, we confine all the projection measurements in the X-Z plane of the Bloch Sphere and assume $a_2=a_3=a_4=a$, $b_2=b_3=b_4=b$ due to the symmetry of the inequality. The projection angles of the operators are
\begin{eqnarray}
\label{eq:angle}
\alpha_1=2.52^\circ,\ \alpha=48.47^\circ,\ \beta_1=163.70^\circ,\ \beta=83.30^\circ,\nonumber
\end{eqnarray}
e.g., $a_1=|\alpha_1\rangle\langle\alpha_1|$ and $|\alpha_1\rangle=\mathrm{cos}\alpha_1|0\rangle+\mathrm{sin}\alpha_1|1\rangle$.
Using such measurement settings, we can calculate that the fidelity of the four-photon GHZ state should be higher than $95\%$ in order to violate the inequality, assuming the noise to be white noise. The measurement result is shown in Table I. The overall data collection time is over 40 hours. Our result shows a violation of the inequality by more than 4 standard deviations, which demonstrates the genuine multipartite nonlocality of the observed state using only 8 joint measurement settings.

\section{Conclusion}

There are three main imperfections in our system which limit the quality of the observed GHZ state. One is the imperfection of our EPR source, the fidelity of which is about 0.99. The second is the higher-order photon pairs emission noise in the SPDC process, which can be characterized by the measurement in the H/V basis (Fig. 3(a)), because this measurement has nothing to do with the interference visibilities in our system, either the HOM interference or the biphoton entanglement interference. By using a rather low pump power (20 mW) to generate the GHZ state, the error rate of the H/V measurement is reduced to $1\%$. Note that all polarization elements in our system (including PBSs, wave plates, polarization-maintaining of the fibers, etc.) have rather high extinction ratios $(>1000:1)$, which is negligible comparing with other noises. The last noise is the distinguishability between two independent photons, especially the temporal mismatch induced by the pump pulse duration. As depicted in \cite{zukowski1995entangling}, the time jitter of the interfering photons should be much shorter than their coherence length, which can be easily understood because the larger the overlapped area of the photon wave packets is, the higher interference visibility we will observe. In order to increase the visibility, on one hand, we use narrow-band filters (2 nm) to stretch the coherence length of the interfering photons; on the other hand, we finely adjust the mode-locked laser to minimize the pump pulse duration to 90 fs. With these measures, we observe a four-photon HOM interference with a visibility of $97.5\%$.

In summary, we experimentally generate an ultrahigh-fidelity four-photon GHZ state, and demonstrate a high performance entanglement swapping operation with the same setup. We further demonstrate the genuine multipartite nonlocality of the GHZ state in the nonsignaling senario, by testing a hard-to-violate Hardy-like inequality. These results greatly reduce the large fidelity-gap between previous biphoton and multiphoton experiments. Our setup would also be useful in implementing other multi-particle quantum information tasks with high quality.

\bigskip

\noindent \textbf{Funding}

\bigskip

\noindent National Natural Science Foundation of China (No. 61327901, No. 61490711, No. 11274289, No. 11325419, No. 61225025, No. 11474268, No. 11374288, No. 11504253); Strategic Priority Research Program (B) of the Chinese Academy of Sciences (Grant No. XDB01030300); National Youth Top Talent Support Program of National High-level Personnel of Special Support Program (No. BB2470000005); Fundamental Research Funds for the Central Universities (No. WK2470000018, No. WK2470000022).

\bigskip

\noindent \textbf{Appendix}

\bigskip

\noindent Here we show how to build up the sandwichlike EPR source. This new EPR source has a sandwichlike structure: two beamlike-phase-matching type-II BBO crystals with a true-zero-order half-wave plate (THWP) between them. The two BBO crystals have the same cutting angles for beamlike SPDC emission and are placed in the same way, so they both produce the product state $|H\rangle|V\rangle$. The THWP in the middle works at the wavelength of the down-conversion photons and is rotated to $45^\circ$, so the polarization states of the idler and signal photons from the first BBO are exchanged through it. When the two possible ways of generating photon pairs (through the first or the second BBO) are further made indistinguishable by spatial and temporal compensations, the photon pairs are prepared in the polarization entangled state $\frac{1}{\sqrt{2}}\left(|H\rangle|V\rangle-|V\rangle|H\rangle\right)$. The main advantage of the sandwichlike structure compared to the previous beamlike source \cite{Niu:08} is that the photons in the same path have the same ordinary (extraordinary) spectral property. Thus it has the same spirit with the ``entanglement concentration" scheme \cite{PhysRevA.67.010301}, which can erase the additional timing information presented in the ultrafast pump pulse and lead to an entanglement source independent of spectral filtering, crystal thickness, and pump bandwidth. Furthermore, our source doesn't need the HOM interferometer, which is critical in the original concentration scheme.

The experimental setup of the sandwichlike source is shown in Fig. 1. The pump laser comes from the second-harmonic generation (SHG) of a mode-locked Ti:Sapphire laser (with a duration of 140 fs, a repetition rate of 76MHz, and a central wavelength of 780 nm), which is not shown in Fig. 1, for simplicity. The SHG is done by focusing the 780 nm Ti:Sapphire laser beam onto a 1.5-mm-thick type-I collinear phase-matching BBO crystal, with a 150-mm-focal-length plano-convex lens. The generated 390 nm laser beam is then collimated by a plano-convex lens with 75 mm focal length. The lens L1 of 150 mm focal length is used to focus the collimated pump beam onto the sandwichlike strategy for down-conversion. The two BBO crystals are both 1-mm-thick and cut at $\theta=42.62^\circ$, $\phi=30^\circ$. The THWP is inserted between the two BBOs, and sticked with them, using very thin (0.05 mm$\pm$0.02 mm thick) and narrow glass stripes and optic glue between them. The glass stripes (together with optic glue) are placed along the edges of the BBO, i.e., outside their clear aperture, to avoid optical damage by strong ultraviolet pump laser. Finally, the gap between the THWP and each BBO is estimated to be in the range from 0.04 mm to 0.08 mm, due to the uncertain thickness of the optical glue. The two L2 lenses of 125 mm focal length are used to collimate the down-converted beams. Note that the optimal collimating positions of the L2 lenses are not the same for the two BBO crystals due to the gap between them. To make the gap as small as possible we use the THWP and bound the three crystals together with very thin glass stripes, which can benefit the collection efficiency. Even with these measures, the difference positions for L2 still exist and in general we choose the intermediate position, which will lead to a $2\%\sim3\%$ decrease of the collection efficiency for both BBO crystals. The down-converted photons are finally coupled into single-mode fibers with aspheric lenses (Thorlabs, model: F220FC-B). The coupling efficiency is about $2/3$ in our system, and the total photon pair collection efficiency is $29\%$, when 3 nm and 8 nm FWHM filters are used for e- and o-photons respectively. The different filtering bandwidths for e- and o-photons are due to the asymmetric generation bandwidth of them. The 3 nm bandwidth is the usually chosen bandwidth for multi-photon interferences. And we determine the 8 nm bandwidth by combining two different bandpass filters (Semrock, LD01-785/10 and TBP01-800/12) and tilting them to form a bandwidth-tunable filter, then watching the corresponding collection efficiency to get a best value.

Compared with the traditional type-I and type-II entanglement sources, the only additional requirement of our source is the spatial compensation, which is realized by two LiNbO$_3$ crystals. They are cut with the optic axes lying in the horizontal plane and being $45^\circ$ away from the incident beam, thus working as a beam-displacer. The thicknesses of them are 1 mm and 3.2 mm for o- and e-photons, respectively. To check whether the compensation is accomplished, we can translate the fiber coupler (together with the fiber) along the horizontal direction perpendicular to the incident down-converted beams, to see whether there is only one peak value of the collected photon counting rates. We can also use this method to determine the thickness of the LiNbO3 crystal, because the translation distance is proportional to the thickness of the crystal when the lenses are removed. Note that the LiNbO$_3$ crystals should be placed before the collimating lenses L2. We use YVO$_4$ crystals for temporal compensations. The thicknesses are 0.6 mm and 0.42 mm for o- and e-photons respectively. In practice, due to the standard errors of thicknesses and distances in manufacturing the sandwichlike structures, the thicknesses of the YVO$_4$ crystals may be slightly different in different cases. But the biggest differences we observed are only about 0.03 mm, which corresponds to a $8\lambda$-order quartz compensation plate ($\lambda=$780 nm).

\end{document}